%% file: skeleton.tex
\documentclass[a4paper,11pt]{article}
\usepackage{pos}
\usepackage{bm}
\usepackage[nolist, nohyperlinks]{acronym}
\usepackage{natbib}

\title{Spectra for the Vacuum Cherenkov Effect in Astrophysical Electromagnetic Cascades with Lorentz Invariance Violation}
\ShortTitle{Spectra for Vacuum Cherenkov in Astrophysical Electromagnetic Cascades with LIV}

\author*[a,b,c]{Andrey Saveliev}
\author[d,e]{Rafael {Alves Batista}}
\author[f]{Feodor Mishin}

\affiliation[a]{\footnotesize Immanuel Kant Baltic Federal University, Institute of High Technology, Kaliningrad, Russia}

\affiliation[b]{\footnotesize I. Kant Baltic Federal University, S.~Kovalevskaya North-West Center for Math.~Research, Kaliningrad, Russia}

\affiliation[c]{\footnotesize Lomonosov Moscow State University, Faculty of Comput.~Math.~and Cybernetics, Moscow, Russia}

\affiliation[d]{\footnotesize Sorbonne Universit\'e, Institut d’Astrophysique de Paris (IAP),
CNRS UMR 7095, Paris, France}

\affiliation[e]{\footnotesize Sorbonne Universit\'e, Laboratoire de Physique Nucléaire et de Hautes Energies (LPNHE), Paris, France}

\affiliation[f]{\footnotesize Harrow School, Harrow, United Kingdom}

\emailAdd{anvsavelev@kantiana.ru}

\abstract{Lorentz invariance violation is a feature of several quantum gravity models in which Lorentz symmetry is broken at high energies, possibly leading to changes in particle behavior and interactions. In this work, we investigate vacuum Cherenkov radiation, a reaction in which an electron spontaneously emits a photon. This process, forbidden when  Lorentz symmetry is unbroken, is a phenomenological consequence of some quantum gravity models. We derive, for the first time, the spectra for the vacuum Cherenkov reaction, and confirm our results numerically. These results can be used to derive limits on Lorentz invariance violation.}

\ConferenceLogo{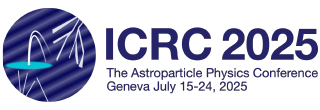}

\FullConference{39th International Cosmic Ray Conference (ICRC2025)\\
 15–24 July 2025\\
Geneva, Switzerland\\}

\begin{document}
\maketitle

\input{acronyms}

\section{Introduction}

The \ac{SM} successfully describes electromagnetic, weak, and strong interactions but leaves key questions unanswered, such as the nature of dark matter and dark energy, the origin of neutrino masses, and the hierarchy problem, while also excluding gravity. Uniting the \ac{QFT} of the \ac{SM} with \ac{GR} has led to the pursuit of a theory of \ac{QG}, which attempts to unify the \ac{SM} and \ac{GR}. These effects are presumed to become relevant at
the Planck scale which is far beyond current experimental capabilities like those of the Large Hadron Collider. Nevertheless, certain \ac{QG} effects might still be detectable through phenomena such as particle propagation on cosmic scales or frame-dependent energy shifts~\cite{Addazi:2021xuf}. 

One potential signal of \ac{QG} is \ac{LIV}. A common approach to study \ac{LIV} within field theory is by extending the \ac{SM} Lagrangian with additional terms, resulting in an effective field theory that captures possible deviations from Lorentz symmetry.

In terms of particle dynamics, phenomenologically, \ac{LIV} primarily affects particle propagation by altering the dispersion relation, which then takes the form
\begin{equation}
E_{\rm LIV}^{2} = E_{\rm SM}^{2} + f_{\rm LIV}(p)\,,
\label{eq:generalDR}
\end{equation}
where $E_{\rm LIV}$ is the energy of the particle in the presence of \ac{LIV}, $E_{\rm SM}$ is its energy without \ac{LIV} and $f_{\rm LIV}(p)$ is the shift due to \ac{LIV},  usually dominated by a single power of the particle's momentum $p$, i.e.~$f_{\rm LIV}(p) \propto \mathcal{O}(p^{n+2})$ with $n \geq 0$. This modification impacts the reaction thresholds \cite{Jacobson:2002hd,JCAP10-03-046} and, consequently, alters the propagation length of particles.

By allowing Lorentz symmetry to be broken, \ac{LIV} leads to a wealth of new processes that would otherwise be forbidden, such as spontaneous photon decay, photon splitting, \ac{VC} radiation from charged particles, and even spontaneous disintegration of atomic nuclei~\cite{alvesbatista2025a}. This ultimately impacts how particles travel in the universe.

In this work, we compute the spectra of photons resulting from \ac{VC} radiation emitted during the propagation of \acl{HE} electrons and positrons in astrophysical environments~\cite{Kaufhold:2005vj, Anselmi:2011ae, Schreck:2017isa}.

\section{The kinematics of the vacuum Cherenkov effect}

We consider modified dispersion relations that take the general form
\begin{equation}
\label{EDisp}
E_{e}^{2} = m_{e}^{2} + p_{e}^{2} + \sum\limits_{n=0}^{\infty} \chi_{n}^{e} \dfrac{p_{e}^{n+2}}{M_{\rm Pl}^{n}} \;,\;\;\; E_{\gamma}^{2} = k_{\gamma}^{2} + \sum\limits_{n=0}^{\infty}\chi_{n}^{\gamma} \dfrac{k_{\gamma}^{n+2}}{M_{\rm Pl}^{n}} 
\end{equation}
for electrons/positrons and photons, respectively.
We closely follow the formalism from Ref.~\cite{Rubtsov:2012kb}, restricting ourselves to the second-order case ($n=2$), which features in a subset of the minimal~\cite{Colladay:1998fq} and non-minimal~\cite{Kostelecky:2009zp,Kostelecky:2013rta} \ac{SME}. 

\begin{figure}[ht]
    \centering
    \includegraphics[width=0.6\columnwidth]{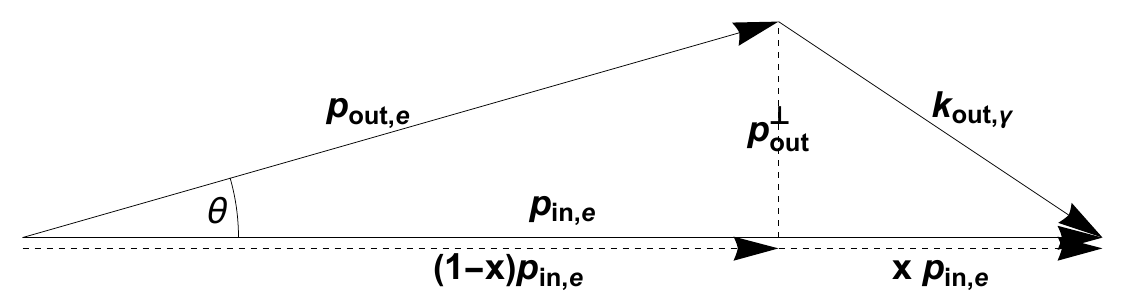}
    \caption{\ac{VC} kinetics formalism used in \cite{Rubtsov:2012kb} and in the present work for the momenta of the incoming electron, outgoing photon and outgoing electron labeled $p_{{\rm in},e}$, $k_{{\rm out},\gamma}$ and $p_{{\rm out},e}$, respectively.}
    \label{fig:VCkinematics}
\end{figure}

For the formalism shown in Fig.~\ref{fig:VCkinematics} conservation of energy and momentum implies~\cite{Rubtsov:2012kb}
\begin{equation} \label{omegapperp}
\frac{p_{\rm out,\perp}^{2}}{2 p_{{\rm in},e} x (1-x)} = \omega_{\rm LV}^{\rm VC}(x) \equiv 
- \frac{\chi_{2}^{\gamma}}{2} \frac{p_{{\rm in},e}^{3} x^{3}}{M_{\rm Pl}^{2}} + \frac{\chi_{2}^{e}}{2} \frac{p_{{\rm in},e}^{3} \left(x^{3} - 3 x^{2} + 3 x\right)}{M_{\rm Pl}^{2}}\,,
\end{equation}
which can be promptly used to compute the \ac{VC} differential rate ${\rm d}\Gamma_{\rm VC}/{\rm d}x$ in terms of $x$ \cite{Rubtsov:2012kb},
\begin{equation} 
\dfrac{{\rm d}\Gamma_{\rm VC}}{{\rm d}x} = \alpha \left( \dfrac{2}{x} - 2 + x \right) \omega_{\rm LV}^{\rm VC}(x)\,,
\label{dGammadxVC}
\end{equation}
with $\alpha$ being the fine-structure constant.

\section{Probabilities and Rates for the Vacuum Cherenkov Effect}

This work is part of a larger ongoing program aiming to perform detailed Monte Carlo simulations of electromagnetic cascades considering \ac{LIV}. For this reason, we compute the detailed spectrum of the emitted photons, whose total emission rate ($\Gamma_{\rm VC}$) can be written as\footnote{For a full derivation, the reader is referred to Ref.~\cite{Saveliev:2024whq}.}:
\begin{equation} 
\Gamma_{\rm VC} = 
\begin{cases}
\Gamma_1 \equiv \alpha \mathcal{G}_{0} \dfrac{p_{{\rm in},e}^{3}}{M_{\rm Pl}^{2}}  \,, & x \in (0;1)\,, (\chi_{2}^{e} \ge 0) \wedge (\chi_{2}^{\gamma} \le \chi_{2}^{e})\,, \\
\Gamma_2 \equiv \alpha \mathcal{G}_{+} \dfrac{p_{{\rm in},e}^{3}}{M_{\rm Pl}^{2}}  \,, & x \in (0;x_{\rm VC,+})\,, 0 < \chi_{2}^{e} < \chi_{2}^{\gamma}\,, \\
\Gamma_3 \equiv \alpha (\mathcal{G}_{0} - \mathcal{G}_{-})  \dfrac{p_{{\rm in},e}^{3}}{M_{\rm Pl}^{2}} \,, & x \in (x_{\rm VC,+};1)\,, \chi_{2}^{\gamma} < \chi_{2}^{e} < 0\,, \\
[\text{no VC}]\,, & \text{else,}
\end{cases}
\label{GammaVCcases}
\end{equation}
where
\begin{equation} \label{omegaVCeq0}
x_{\rm VC,\pm} = - \frac{3 \chi_{2}^{e}}{2\left(\chi_{2}^{\gamma} - \chi_{2}^{e} \right)} \pm \frac{\sqrt{3 \chi_{2}^{\rm e} \left(4\chi_{2}^{\gamma} - \chi_{2}^{e} \right)}}{2 \left|\chi_{2}^{\gamma} - \chi_{2}^{e} \right|}\,.
\end{equation}

A final point to be considered concerns the kinematical threshold for the reaction; \ac{VC} is only possible if the momentum of the incoming electron exceeds a threshold value $p_{\rm VC,thr}$, i.e.~$p_{\rm in,e} > p_{\rm VC,thr}$. For the case considered here, \ac{LIV} of order $n=2$, the only non-zero coefficients are $\chi_{2}^{e}$ and $\chi_{2}^{\gamma}$. In this case the threshold values may be found in Ref.~\cite{Jacobson:2002hd}.

The existence of a (lower) threshold, together with the integration limits for $x$, may be used in specific cases to determine the momenta ($k_{\rm out,\gamma}$ and $p_{{\rm out},e}$) of the outgoing particles.

With these results we now can calculate the total interaction rate for \ac{VC}, $\Gamma_{\rm VC}$, by integrating Eq.~(\ref{dGammadxVC}) in the range specified in Eq.~(\ref{GammaVCcases}), for a given combination of $\chi_{2}^{e}$ and $\chi_{2}^{\gamma}$, obtaining
\begin{equation} 
\Gamma_{\rm VC} = 
\begin{cases}
\Gamma_1 \equiv \alpha \mathcal{G}_{0} \dfrac{p_{{\rm in},e}^{3}}{M_{\rm Pl}^{2}}  \,, & x \in (0;1)\,, (\chi_{2}^{e} \ge 0) \wedge (\chi_{2}^{\gamma} \le \chi_{2}^{e})\,, \\
\Gamma_2 \equiv \alpha \mathcal{G}_{+} \dfrac{p_{{\rm in},e}^{3}}{M_{\rm Pl}^{2}}  \,, & x \in (0;x_{\rm VC,+})\,, 0 < \chi_{2}^{e} < \chi_{2}^{\gamma}\,, \\
\Gamma_3 \equiv \alpha (\mathcal{G}_{0} - \mathcal{G}_{-})  \dfrac{p_{{\rm in},e}^{3}}{M_{\rm Pl}^{2}} \,, & x \in (x_{\rm VC,+};1)\,, \chi_{2}^{\gamma} < \chi_{2}^{e} < 0\,, \\
[\text{no VC}]\,, & \text{else}
\end{cases}
\label{GammaVCcases}
\end{equation}
where
\begin{equation} \label{omegaVCeq0}
x_{\rm VC,\pm} = - \frac{3 \chi_{2}^{e}}{2\left(\chi_{2}^{\gamma} - \chi_{2}^{e} \right)} \pm \frac{\sqrt{3 \chi_{2}^{\rm e} \left(4\chi_{2}^{\gamma} - \chi_{2}^{e} \right)}}{2 \left|\chi_{2}^{\gamma} - \chi_{2}^{e} \right|}\,,
\end{equation}
and the parameters $\mathcal{G}_{\pm}$ are defined as
\begin{equation}
\begin{split}
& \mathcal{G_{\pm}} \equiv 
 \left[ 37 \left( \pm \mathcal{S} - 6 \chi_{2}^{\gamma} \right) \chi_{2}^{e} \chi_{2}^{\gamma} - 64 \left(\chi_{2}^{e}\right)^{3} - \left( \pm 14 \mathcal{S} - 207 \chi_{2}^{\gamma} \right) \left(\chi_{2}^{e}\right)^{2} - 10 \left( \pm 5 \mathcal{S} - 16 \chi_{2}^{\gamma} \right) \left(\chi_{2}^{\gamma} \right)^{2} \right] \\
&\times \frac{\chi_{2}^{e} \left( \pm S - 3 \chi_{2}^{e} \right)}{160 \left( \chi_{2}^{\gamma} - \chi_{2}^{e} \right)^{4}}\,, \;\;\; \text{and} \;\;
\mathcal{G}_{0} \equiv \dfrac{157 \chi_{2}^{e} - 22 \chi_{2}^{\gamma}}{120} \,,
\end{split}
\end{equation}
wherein $\mathcal{S} \equiv \sqrt{3 \chi_{2}^{\rm e} \left(4\chi_{2}^{\gamma} - \chi_{2}^{e} \right)}$. Employing these definitions, we can now understand the behavior of the \ac{VC} rate considering the $\chi_{2}^{e}$--$\chi_{2}^{\gamma}$ parameter space. The possible values of $\Gamma_{\rm VC}$ for all the different parameter combinations are presented in Eq.~(\ref{GammaVCcases}).

From Eq.~(\ref{GammaVCcases}) we can finally calculate the differential probability ($\text{d}P_\text{VC} / \text{d}x$) with respect to the fraction $x$ of momentum carried away by the photon, which together with the threshold values gives
\begin{equation} \label{dPVCdxcases}
\dfrac{{\rm d}P_{\rm VC}}{{\rm d}x} = \max\left\{ 0, \,
  \alpha \left( \dfrac{2}{x} - 2 + x \right) \dfrac{\omega_{\rm LV}^{\rm VC}(x)}{\Gamma_{\rm VC}} \right\} \,,
\end{equation}
for $p_{{\rm in},e} > p_{\rm VC,thr}$ with $\Gamma_{\rm VC}$ given by Eq.~(\ref{GammaVCcases}). 

Based on the symmetry of the allowed values of $x$, cf.~Eq.~(\ref{GammaVCcases}), we illustrate Eq.~(\ref{dPVCdxcases}) by exemplarily considering parameter combinations which obey the relation $\left| \chi_{2}^{\gamma}  \chi_{2}^{e} \right| = 10^{-10}$.

\section{Results}

We present the \ac{VC} spectra for incoming electrons with initial energy $E_{{\rm in},e} = 10^{21}\,{\rm eV}$, exploring various combinations of \ac{LIV} parameters. For each scenario, we generate $10^{6}$ random samples following Eq.~(\ref{dPVCdxcases}). These simulations account for the full \ac{VC} cascade, i.e.~the repeated \ac{VC}-induced photon emission until the electron energy drops below the \ac{VC} threshold.

The emission of \ac{VC} photons is treated as instantaneous, as the associated emission rate ($\Gamma_{\rm VC}$) implies distances much shorter than all other relevant processes acting on galactic and cosmological scales. Additionally, we employ the co-linear approximation, wherein the emitted particles are parallel to their parent -- a good approximation for high-energy particles.

\begin{figure*}[htb!]
    \centering
    \includegraphics[width=0.45\columnwidth]{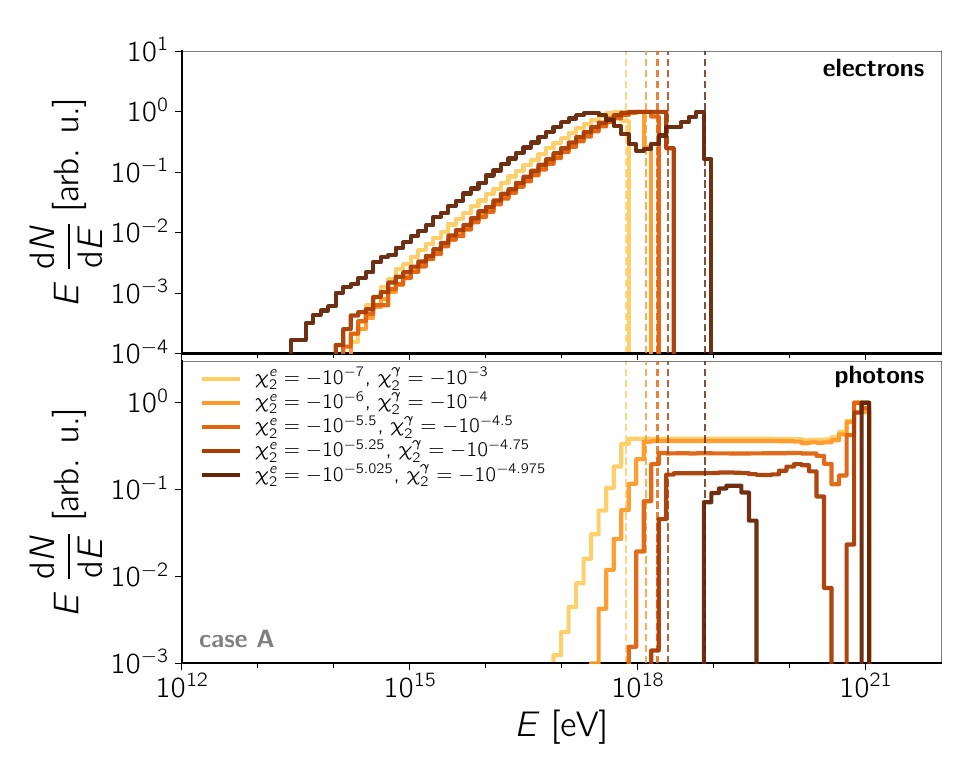}
    \includegraphics[width=0.45\columnwidth]{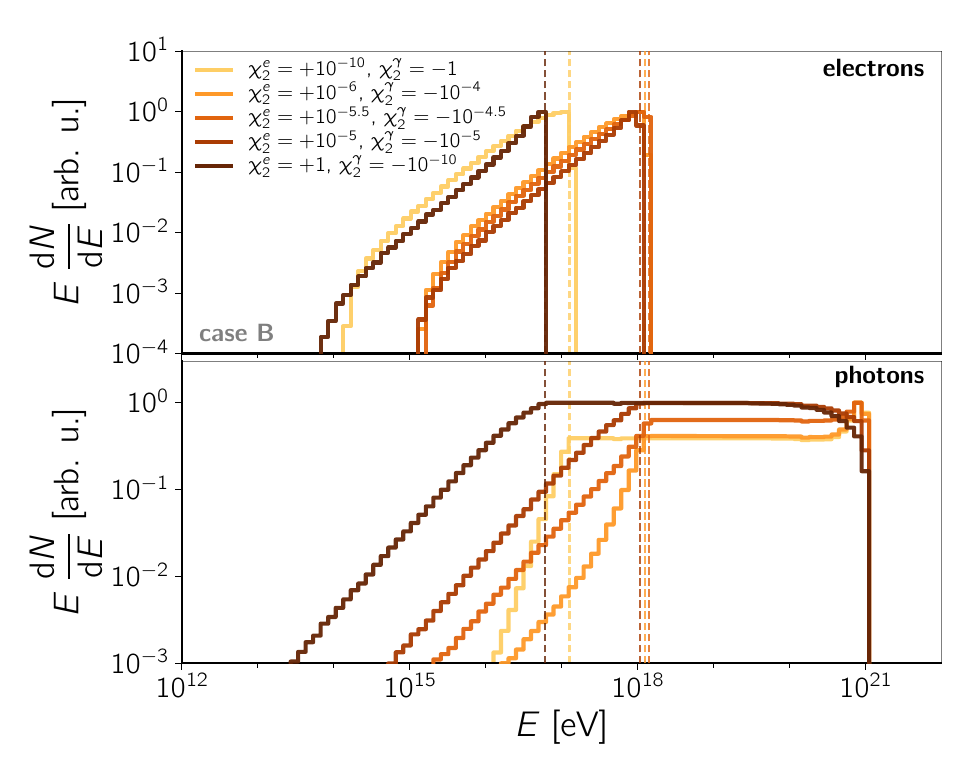}\\[-8pt]
    \includegraphics[width=0.45\columnwidth]{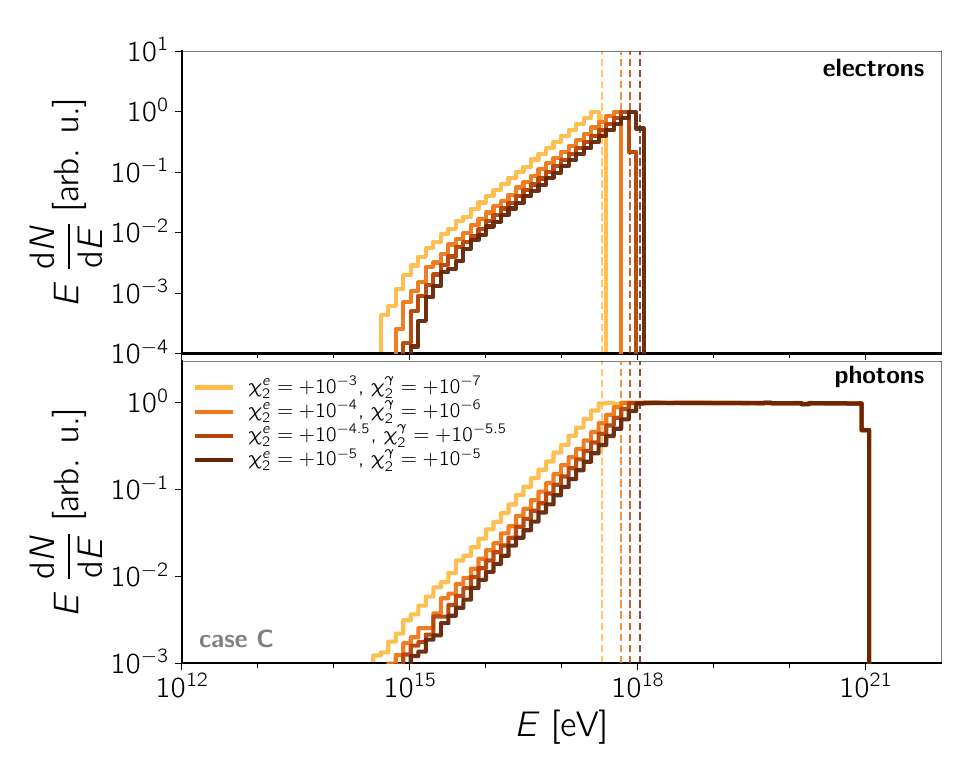}
    \includegraphics[width=0.45\columnwidth]{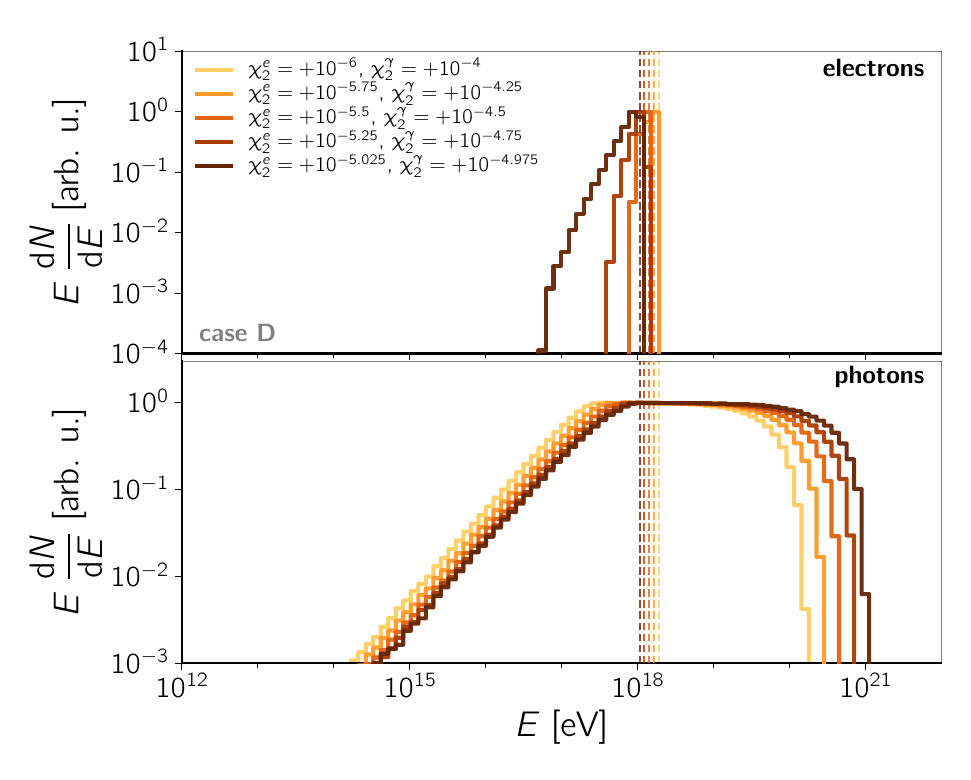}\\[-8pt]
    \caption{Results of Monte Carlo simulations for various $\chi_2^e$ and $\chi_2^\gamma$ combinations. Dashed lines refer to the corresponding threshold energies. Spectra are energy-weighted to highlight the ${\rm d}N/{\rm d}E \propto E^{-1}$ behaviour. 
    The normalization of the distributions is arbitrary.}
    \label{fig:sim_a}
\end{figure*}

We identify four distinct parameter combinations, each corresponding to qualitatively different emission spectra. We refer to them as cases A--D, defined as follows:
\begin{equation}
{\rm A}: \chi_{2}^{\gamma} < \chi_{2}^{e} < 0\,,\;\;  
{\rm B}: \chi_{2}^{\gamma} < 0 \le \chi_{2}^{e}\,,\;\;
{\rm C}: 0 < \chi_{2}^{\gamma} \le \chi_{2}^{e} \,,\;\;
{\rm D}: 0 < \chi_{2}^{e} < \chi_{2}^{\gamma} \,.
\label{eq:cases}
\end{equation}
An important finding emerging directly from Fig.~\ref{fig:sim_a} is the appearance of a lower-energy cut-off in the spectrum of one of the outgoing particles, depending on the parameter regime. 
This is evident for the electron spectra corresponding to \ac{LIV} parameters within range~D, and in the photon spectra for parameters within range~A.

This behaviour can be understood by examining the integration limits. For $x \in (0, x_{\rm VC,+})$, corresponding to the condition $0 < \chi_{2}^{e} < \chi_{2}^{\gamma}$ as defined in Eq.~(\ref{GammaVCcases}), we have $x < x_{\rm VC,+}$, implying  $1 - x > 1 - x_{\rm VC,+}$. Since the minimum incoming electron momentum is set by the \ac{VC} threshold ($p_{\rm VC,thr}$), the outgoing electron momentum satisfies
$p_{\rm VC,min} \le p_{{\rm out},e} \le p_{\rm VC,thr}$ for $0 < \chi_{2}^{e} < \chi_{2}^{\gamma}$, where $p_{\rm VC,min}$ is defined as $p_{\rm VC,min} \equiv (1-x_{\rm VC,+}) p_{\rm VC,thr}$.
Conversely, for the integration limits $x \in (x_{\rm VC,+};1)$, corresponding to  $\chi_{2}^{e} < \chi_{2}^{\gamma} < 0$, we have $x \ge x_{\rm VC,+}$ and hence $1-x < 1-x_{\rm VC,+}$. This same reasoning also implies that $k_{{\rm out},\gamma} \ge k_{\rm VC,min}$ for $\chi_{2}^{\gamma} < \chi_{2}^{e} < 0$, where $k_{\rm VC,min}$ is defined as $k_{\rm VC,min} \equiv x_{\rm VC,+} p_{\rm VC,thr}$.

\begin{figure*}[htb!]
    \centering
    \includegraphics[width=0.45\columnwidth]{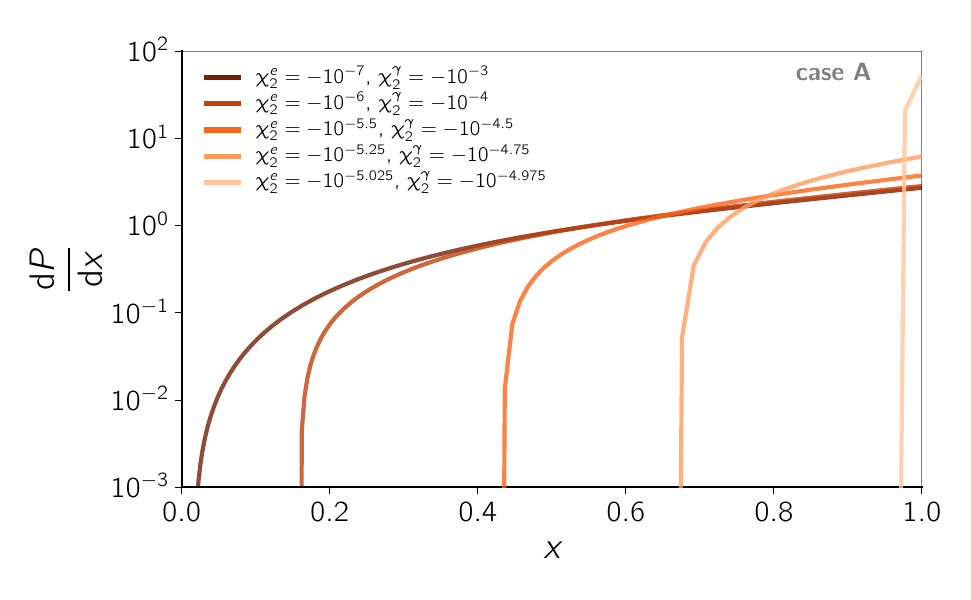}
    \includegraphics[width=0.45\columnwidth]{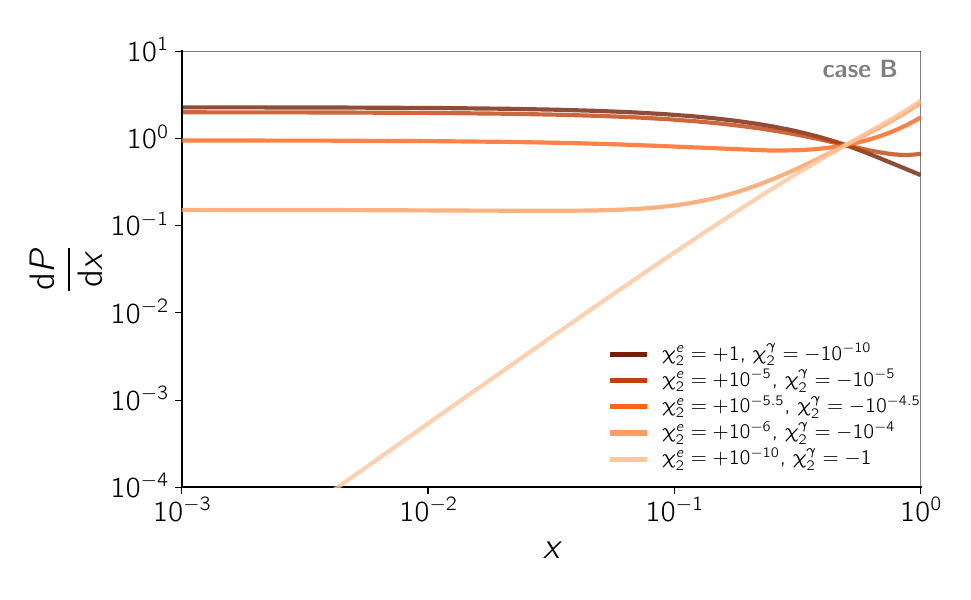} \\[-8pt]
    \includegraphics[width=0.45\columnwidth]{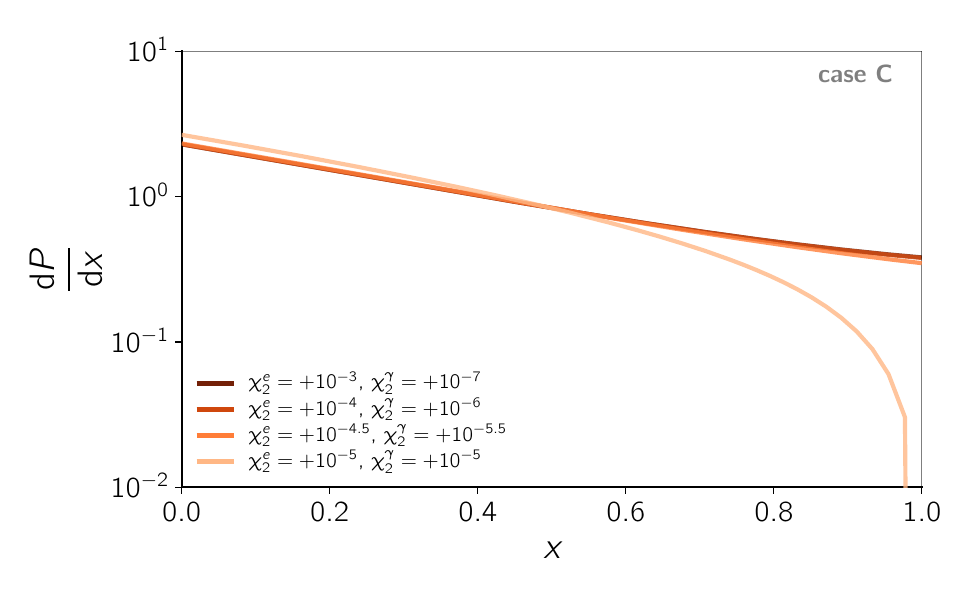}
    \includegraphics[width=0.45\columnwidth]{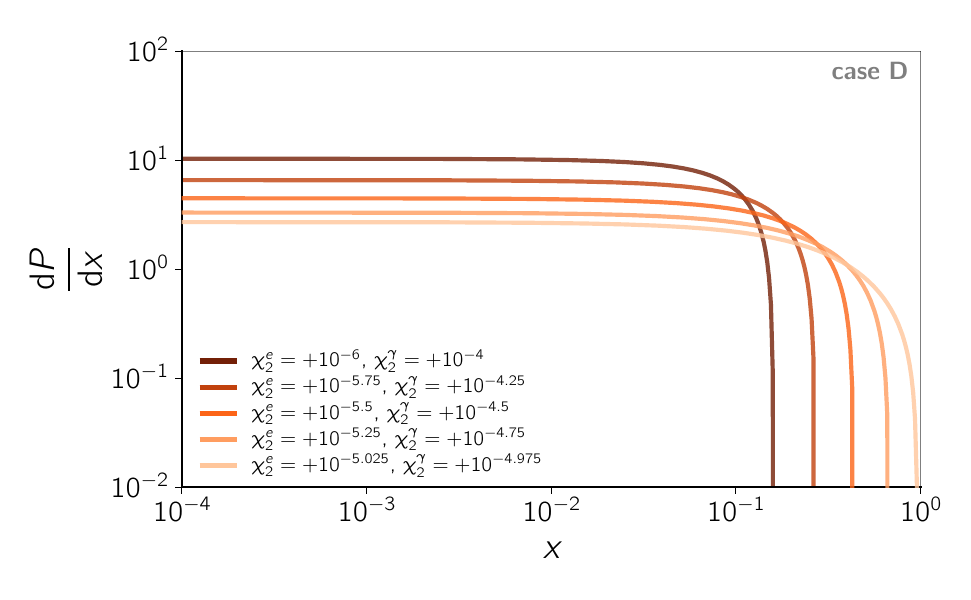} \\[-8pt]
    \caption{Differential probability distribution ($\frac{{\rm d}P_{\rm VC}}{{\rm d}x}$) for various combinations of $\chi_{2}^{e}$ and $\chi_{2}^{\gamma}$, with $\left| \chi_{2}^{\gamma}  \chi_{2}^{e} \right| = 10^{-10} $. Each panel corresponds to different regimes, based on the signs of $\chi_{2}^{\gamma}$ and $\chi_{2}^{e}$, according to Eq.~(\ref{eq:cases})}.
    \label{fig:dPdx}
\end{figure*}

Another prominent feature common to all photon spectra is that all of them follow a particular power law, extending from the \ac{VC} threshold up to nearly the initial electron energy (see Fig.\ref{fig:sim_a}). This arises because the differential emission probability is independent of the incoming electron momentum, as given in Eq.(\ref{dPVCdxcases}).

We now turn to the different cases of the photon spectra, starting with a feature common to the cases C and D. While for higher energies one can see the power-law behavior ${\rm d}N/{\rm d}E_{{\rm out},\gamma}$ of proportional to $E_{{\rm out},\gamma}^{-1}$ as described above, the spectrum flattens to ${\rm d}N/{\rm d}E_{{\rm out},\gamma} \propto E_{{\rm out},\gamma}^{0}$ for lower energies. This is an immediate consequence of the differential probability distribution (see Eq.~(\ref{dPVCdxcases})) being peaked around zero and then falling roughly linearly for higher values of $x$, as shown in the corresponding panels of Fig.~\ref{fig:dPdx}. The transition between these two regimes is determined by the \ac{VC} threshold value.

The photon spectra for case A exhibits a lower cut-off and, for $\chi_{2}^{\gamma}$ and $\chi_{2}^{e}$ being close to each other, a gap near the initial momentum, which widens as the two values converge. This gap-like feature is once more a direct consequence of the corresponding differential probability distributions (cf.~Fig.~\ref{fig:dPdx}). Here we can see that the closer the two values are to each other the narrower the differential distribution, which peaks around $1$, becomes, producing a narrow peak in the photon spectrum close to the initial momentum value and then another peak before the electron drops below the threshold value. Once the values get further apart, the distribution widens, which itself results in a widening of the peak as well as its shift to lower momentum values.

Finally, for case B there are two different regimes for ${\rm d}N/{\rm d}E_{{\rm out},\gamma}$ at lower momentum values of the outgoing photons -- either a quadratic increase for $\left| \chi_{2}^{\gamma} \right| \gg \chi_{2}^{e}$ or a flat spectrum for $\left| \chi_{2}^{\gamma} \right| \ll \chi_{2}^{e}$. This behavior may again be explained by analyzing the corresponding plots in Fig.~\ref{fig:dPdx}. Both regimes correspond directly to the behavior of the corresponding differential probability densities, as due to the fact that they are independent of the momentum of the incoming electron, the spectrum is effectively a superposition of many of such individual distributions.

Proceeding to the electron spectra, for case A we see that the spectral shape of ${\rm d}N/{\rm d}E_{{\rm out},e}$ is a flat spectrum with a sharp cut-off which occurs when the electron momentum drops below the \ac{VC} threshold value. The only spectral variation appears right below the cut-off where the spectrum either slightly rises or falls. This is, again, a direct consequence of the corresponding differential probability distribution function, but this time resulting in a reverse relationship compared to the photon spectra which is due to the fact that the electron carries away the remaining fraction of the incoming momentum, $1-x$.

The other case for which the electron spectrum displays some interesting features (apart from the lower cut-off described above) is for \ac{LIV} parameter values lying within the ranges denoted as case D. If $\chi_{2}^{e}$ and $\chi_{2}^{\gamma}$ are fairly apart from each other, one gets an almost monochromatic emission due to the differential probability shown in the top panel of Fig.~\ref{fig:dPdx}. We can see that for electrons it is tightly peaked around $1$, which translates into the monochromatic spectrum. By further emitting photons, this distribution simply moves to lower momentum values until it reaches the threshold value at which point it does not change anymore. Once the values of $\chi_{2}^{e}$ and $\chi_{2}^{\gamma}$ are getting close to each other, this nearly monochromatic spectrum  acquires a low-momentum tail, again simply resulting directly from the now wider differential probability distribution.

\section{Discussion and Outlook}

In this work, we investigated if the \acl{VC} leaves a discernible imprint in high-energy gamma-ray data. We found that in regions like the Milky Way, where magnetic fields are modest ($\sim 0.1 \; \text{nT}$), that \ac{VC} dominates over synchrotron losses by several orders of magnitude. This is not true in the surroundings of highly magnetized compact objects, where strong magnetic fields incur synchrotron losses that depend quadratically on the field. However, the radiated synchrotron power when Lorentz symmetry is broken may not follow trivially.

Compared to inverse Compton scattering of energetic electrons off photon backgrounds such as the \ac{CMB} and the \ac{EBL}, \ac{VC} is also more efficient.

While our primary focus was on electrons and positrons, our findings can be immediately extended to other charged leptons, namely muons and tauons, provided that their lifetime exceeds the \ac{VC} time scale.

A noteworthy feature of our framework is that we allow the Lorentz-violating coefficients to differ by particle species.  Because \ac{CPT}-odd terms flip the signs between particles and antiparticles, this distinction is indispensable for any realistic phenomenology.

For simplicity, we based our analysis on direction-independent \ac{LIV} terms and worked in a preferred frame. While this makes simulations more manageable, it inevitably masks possible anisotropic signatures that could, in principle, be observed.

We presented the first complete derivation of the \acl{VC} radiation spectrum based on the interaction rates from \cite{Rubtsov:2012kb}. We find that the commonly used ``binary'' approach -- assuming the electron drops to the VC threshold and the photon carries away the momentum difference -- is generally inaccurate. While outgoing electrons can be nearly monochromatic in some cases (see Fig.~\ref{fig:sim_a}), the photon spectrum is typically broader, requiring a full treatment of the \ac{VC} process including successive energy losses.

Due to the short interaction time, \ac{VC} radiation occurs almost instantly after electron creation, making the resulting spectra valuable inputs for modeling electromagnetic cascades over Galactic and larger scales. As part of a larger-term plan~\cite{alvesbatista2025a}, we plan to integrate this feature into the CRPropa framework \cite{alvesbatista2016a, AlvesBatista:2022vem,ICRC2025_CRPropa}, extending our previous work~\cite{Saveliev:2023urg}.

Future studies will also explore how \ac{LIV} affects other key processes like inverse Compton scattering, pair production, and photon decay. In particular, scenarios where both \ac{VC} radiation and photon decay are allowed by specific values of $\chi_{2}^{e}$ and $\chi_{2}^{\gamma}$ merit deeper investigation.

\vspace{-0.5cm}

{\footnotesize
\begin{acknowledgments}

\vspace{-0.4cm}

RAB acknowledges the support of the Agence Nationale de la Recherche (ANR), project ANR-23-CPJ1-0103-01. 
\end{acknowledgments}
\vspace{-0.8cm}
\providecommand{\href}[2]{#2}\begingroup\raggedright\endgroup
}

\end{document}

%% file: acronyms.tex
\acrodef{AGN}{active galactic nucleus}
\acrodefplural{AGN}{active galactic nuclei}
\acrodef{A.U.}{astronomical unit}
\acrodef{BSM}{beyond the Standard Model}
\acrodef{CMB}{cosmic microwave background}
\acrodef{CPT}{charge, parity, and time} 
\acrodef{CRB}{cosmic radio background}
\acrodef{CTA}{Cherenkov Telescope Array}
\acrodef{DSR}{deformed (or doubly) special relativity}
\acrodef{EBL}{extragalactic background light}
\acrodef{GR}{general relativity}
\acrodef{GRB}{gamma-ray burst}
\acrodef{HE}{high-energy}
\acrodef{IACT}{imaging air-Cherenkov telescope}
\acrodef{ICS}{inverse Compton scattering}
\acrodef{IGMF}{intergalactic magnetic field}
\acrodef{LHAASO}{Large High Altitude Air Shower Observatory}
\acrodef{LHC}{Large Hadron Collider}
\acrodef{LIV}{Lorentz invariance violation}
\acrodef{mSME}{minimal Standard-Model extension}
\acrodef{PD}{photon decay}
\acrodef{PP}{pair production}
\acrodef{PS}{photon splitting}
\acrodef{QCD}{quantum chromodynamics}
\acrodef{QED}{quantum electrodynamics}
\acrodef{QFT}{quantum field theory}
\acrodef{QG}{quantum gravity}
\acrodef{SM}{Standard Model}
\acrodef{SME}{Standard-Model extension}
\acrodef{SR}{special relativity}
\acrodef{SWGO}{Southern Wide-field Gamma-ray Observatory}
\acrodef{UHE}{ultra-high-energy}
\acrodef{UHECR}{ultra-high-energy cosmic ray}
\acrodef{VC}{vacuum Cherenkov}
\acrodef{VHE}{very-high-energy}